\documentclass[amsmath,amssymb,10pt,reprint, a4paper,final]{revtex4-1}

\usepackage{graphicx}
\usepackage[breaklinks=true,colorlinks=true,linkcolor=blue,urlcolor=blue,citecolor=blue]{hyperref}
\usepackage{graphicx}
\usepackage{dcolumn}
\usepackage{bm}
\usepackage{threeparttable}
\usepackage{multirow}
\usepackage{adjustbox}
\usepackage{amssymb}
\usepackage{lineno, natbib}

\begin{document}

\title {Experimental and Theoretical Investigation on the Possible Half-metallic Behaviour of Equiatomic Quaternary Heusler Alloys: CoRuMnGe and CoRuVZ (Z = Al, Ga)}

\author{Deepika Rani$^{a}$, Lakhan Bainsla$^{a,b}$, K. G. Suresh$^{a}$ and Aftab Alam$^a$}

\address{$^a$Department of Physics, Indian Institute of Technology Bombay, Powai, Mumbai 400076, Maharashtra, India\\ $^b$ WPI Advanced Institute for Materials Research, Tohoku University, Sendai 980-8577, Japan}

\email{aftab@phy.iitb.ac.in}

\begin{abstract}
		In this report, structural, electronic, magnetic and transport properties of quaternary Heusler alloys CoRuMnGe and CoRuVZ (Z = Al, Ga) are investigated. All the three alloys are found to crystallize in cubic structure. CoRuMnGe exhibits L2$_1$ structure whereas, the other two alloys have  B2-type disorder. For CoRuMnGe and CoRuVGa, the experimental magnetic moments are in close agreement with the theory as well as those predicted by the Slater-Pauling rule, while for CoRuVAl, a relatively large deviation is seen. The reduction in the moment in case of CoRuVAl possibly arises due to the anti-site disorder between Co and  Ru sites as well as V and Al sites. Among these alloys, CoRuMnGe has the highest T$\mathrm{_C}$ of 560 K. Resistivity variation with temperature reflects the half-metallic nature in CoRuMnGe alloy.  CoRuVAl shows metallic character in both paramagnetic and ferromagnetic states, whereas the temperature dependence of resistivity for CoRuVGa is quite unusual. In the last system, $\rho$ vs. T curve shows an anomaly in the form of a maximum and a region of negative temperature coefficient of resistivity (TCR) in the magnetically ordered state. The ab initio calculations predict nearly half-metallic ferromagnetic state with high spin polarization of 91, 89 and 93 \% for CoRuMnGe, CoRuVAl and CoRuVGa respectively. In the case of CoRuMnGe, the XRD analysis reveals that the Co and Ru sites are equally probable. Hence, to investigate the electronic properties of the experimentally observed structure, the Co-Ru swap disordered structures of CoRuMnGe alloy are also simulated and it is found that the disordered structures retain half-metallic nature, high spin polarization with almost same magnetic moment as in the ideal structure. Nearly half-metallic character, high T$\mathrm{_C}$ and high spin polarization make CoRuMnGe alloy promising for room temperature spintronic applications. 

\end{abstract}

\maketitle

\section{Introduction} 
In the last few years, quaternary Heusler alloys received enormous interest due to their wide applications in the field of spintronics. Many of them are reported to show half-metallic behavior and thus have high spin polarization.\cite{doi:10.1063/1.4959093} In half-metallic materials, one of the spin bands exhibits metallic character whereas, the other spin band exhibits a gap at the Fermi level. Magnetic materials with high spin polarization and high Curie temperature are desirable to improve the performance of spintronic devices such as magnetic tunnel junctions,\cite{doi:10.1063/1.3330942,doi:10.1063/1.2354026} spin injectors, spin transistors,\cite{doi:10.1002/pssc.200672894, :/content/journals/10.1049/ip-cds_20045196} and spin valves.\cite{doi:10.1063/1.4821243,1882-0786-7-3-033002} Heusler alloys are potential materials in this field because of their stable structure, high T$\mathrm{_C}$ , tunable electronic properties and high spin polarization. Equiatomic quaternary Heusler alloys X$\mathrm{X^\prime}$YZ (where X, $\mathrm{X^\prime}$ , Y are transition metals and Z is a main group element) crystallize in the space group no. 216 with F$\bar{4}$3m symmetry (Y-type structure with prototype LiMgPdSn).\cite{Y} Among the various reported quaternary Heusler alloys only a few crystallize in ordered Y-type structure. Controlling disorder and defects in this class of materials is still a big challenge for the applications, since disorder greatly affects the spin polarization.\cite{0953-8984-19-31-315215, PhysRevB.74.104405, doi:10.1063/1.4929252} There are a large number of reports on 3d- transition elements based quaternary Heusler alloys, but only a few 4d - based (Ru and Rh based) equiatomic quaternary Heusler alloys have been studied experimentally.\cite{BAINSLA2015631,PhysRevB.96.184404,0953-8984-24-4-046001} It would be interesting to see the effect of replacing one of the 3d- element by a 4d- element in quaternary Heusler alloys. For example, CoFeMnGe alloy was found to have considerable amount of DO$_3$ disorder\cite{doi:10.1063/1.4902831} and replacing Fe by a 4d element (Rh and Ru) is expected to improve the structure, as both CoRuMnGe and CoRhMnGe\cite {PhysRevB.96.184404} are found to crystallize in L2$_1$ structure.\\

We synthesized CoRuMnGe (CRMG), CoRuVAl (CRVA) and CoRuVGa (CRVG) equiatomic quaternary Heusler alloys and investigated their structural, magnetic, electronic and transport properties. X-ray diffraction study reveals that in case of CRMG, 50 \% disorder exists between the Co and Ru sites (i.e. 4c and 4d sites are equally probable for Co and Ru atoms), which reduces its symmetry to L2$_1$ structure. CRVA and CRVG alloys show B2 disorder. The experimental saturation magnetization values are in good agreement with the Slater-Pauling rule for CRMG and CRVG, but CRVA shows a relatively large deviation, which may be attributed to the disorder. This suggest that high spin polarization is not possible in CRVA alloy as Slater-Pauling rule is considered to be a prerequisite for half metallic nature.\cite{PhysRevB.66.174429}  The temperature dependence of electrical resistivity is studied in detail for all the three alloys. In case of CRVG, an unconventional behavior is seen in the form of a maximum and a region with semiconducting behavior in the ferromagnetic state. To investigate the half-metallic behavior in these alloys, electronic structure calculations by ab initio method were performed using Perdew, Burke, and Ernzerhof (PBE) potential. We have also studied the Co-Ru swap disordered structures of CoRuMnGe alloy to get a deeper insight into the effect of disorder on its magnetic and electronic properties. Among these three alloys, CRMG is found to be a potential material for spintronic applications due to its stable structure, high spin polarization and high T$\mathrm{_C}$ value.  

\section{Experimental Details}
The polycrystalline alloys CoRuMnGe and CoRuVZ ( Z = Al, Ga) were prepared by arc melting the stoichiometric amounts of high purity (at least 99.9\% purity) constituent elements in argon atmosphere. A Ti ingot was used as an oxygen getter to further reduce the contamination.  2\%  extra Mn was taken to compensate the weight loss due to Mn evaporation during melting.\cite{PhysRevB.96.184404} For better homogeneity the ingots formed were flipped and melted several times . After melting, the samples were sealed in a quartz tube and annealed for 7 days at 1073 K followed by furnace cooling.\cite{PhysRevApplied.10.054022} Room temperature X-ray diffraction patterns were taken using Cu-K$_\alpha$ radiation with the help of Panalytical X-pert diffractometer. Crystal structure analysis was done using FullProf Suite software. Magnetization isotherms at 5 K were obtained using a vibrating sample magnetometer (VSM) attached to the physical property measurement system (PPMS) (Quantum design) for fields up to 40 kOe. Thermo-magnetic curves in the higher temperature range were obtained using a VSM attached with high temperature oven, in a field of 100 Oe. Electrical resistivity measurements were done using the four probe method in PPMS, applying a  5 mA current.

\section{Computational Details}
\begin{figure*}
	\centering
	\includegraphics[width=0.9\linewidth]{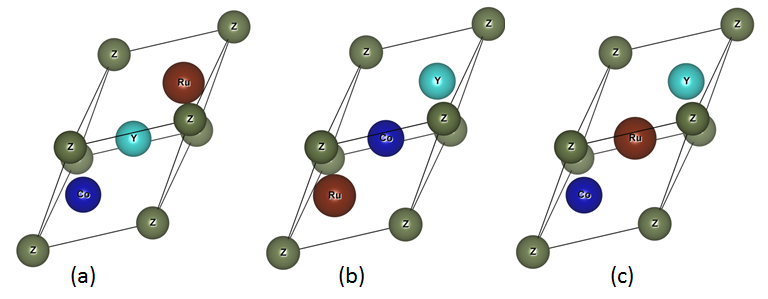}
	\caption{Primitive cell of non degenerate configurations of quaternary Heusler alloys CoRuYZ (Y = Mn, V and Z = Ge, Al, Ga) (a) Type I, (b) Type II and (c) Type III.}
	\label{primitive}
\end{figure*}

Ab initio simulations were performed using a spin resolved density functional theory (DFT) implemented within Vienna ab initio simulation package (VASP) \cite{VASP} with a projected augmented-wave basis.\cite{PAW} The electronic exchange-correlation potential due to Perdew, Burke, and Ernzerhof (PBE) was used within the generalized gradient approximation (GGA) scheme. A $24^3$ k-mesh was used to perform the Brillouin zone integration within the tetrahedron method. A plane wave energy cutoff of 500 eV was used for all the calculations. All the structures are fully relaxed (cell volume, shape, and atomic positions of constituent atoms), with total energies (forces) converged to values less than $10^{-6}$ eV (0.01 eV/\AA). In order to find the most stable crystallographic configuration, four atom primitive cell was used. In general, the possible non degenerate crystallographic configurations for any quaternary Heusler alloy (QHA) XX'YZ, by keeping Z at 4a site, are:
\begin{enumerate}
	\item   X at 4d, X$'$ at 4c and Y at 4b sites  (type I),
	\item	X at 4b , X$'$ at 4d and Y at 4c sites (type II), 
	\item	X at 4d , X$'$ at 4b and Y at 4c  (type III)
\end{enumerate}
Figure \ref{primitive} shows the primitive cells of three distinct configurations of CoRuYZ (Y =Mn, V and Z = Ge, Al, Ga) quaternary Heusler alloys.\\

To study the Co-Ru defects in CoRuMnGe, a $2\times2\times2$ supercell, formed from a four-atom primitive fcc cell of the most stable configuration has been considered. This supercell contains a total of 32 atoms with 8 atoms of each kind. Brillouin zone integrations were performed using $6^3$ k mesh for 32-atoms cell. 

\begin{figure}
\centering
\includegraphics[width=0.7\linewidth]{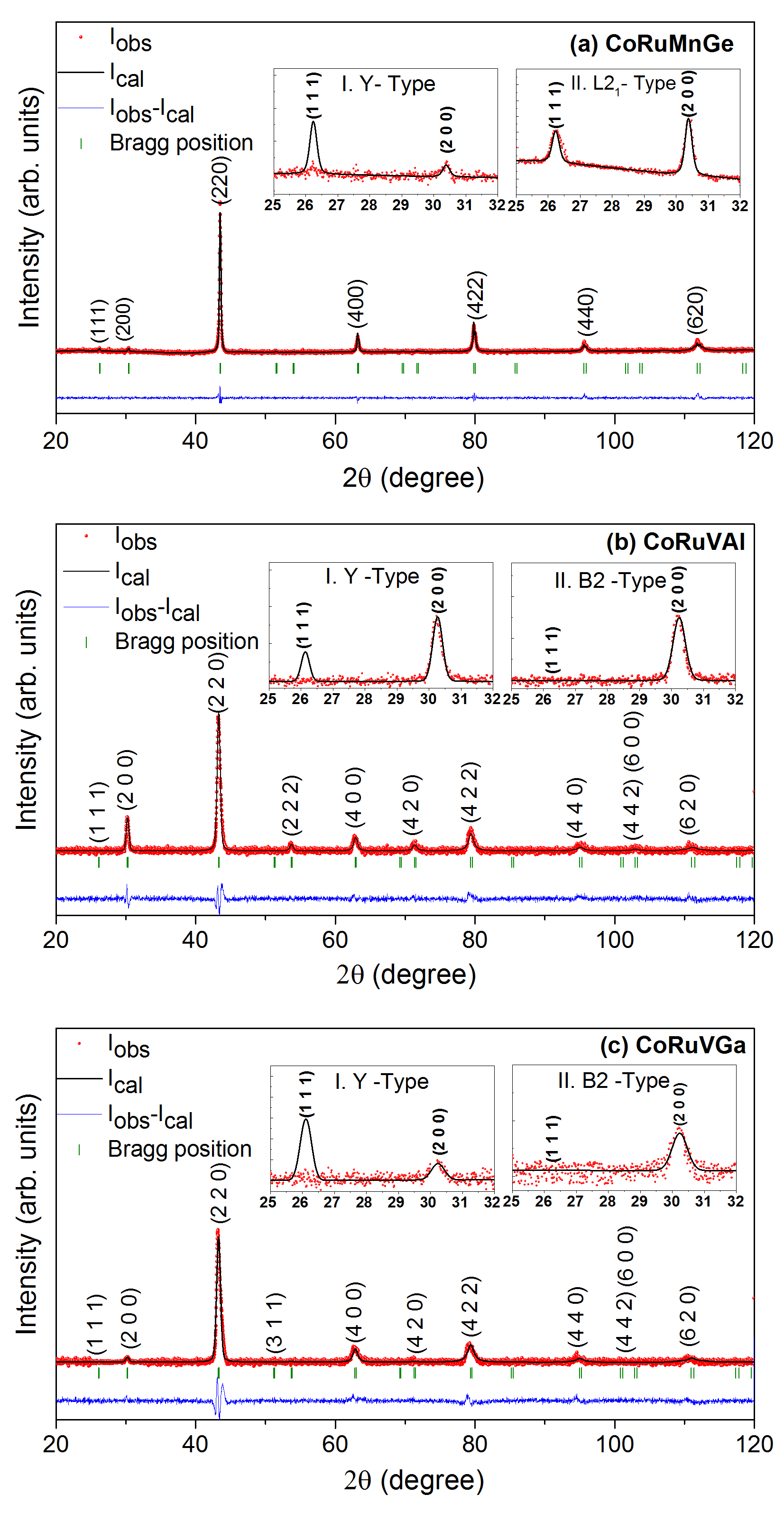}
\caption{Rietveld refined room temperature XRD patterns for (a) CoRuMnGe, (b) CoRuVAl and (c) CoRuVGa alloys. The insets I and II show the observed and calculated intensities of the superlattice reflections (111) and (200) with perfectly ordered LiMgPdSn-type and disordered ($\mathrm{L2_1}$ in case of CRMG and B2 in case of CRVG and CRVA) structure respectively.}
\label{XRD}
\end{figure}

\section{Results and Discussion}
\subsection{Structural analysis}

\begin{figure}
	\centering
	\includegraphics[width=0.5\linewidth]{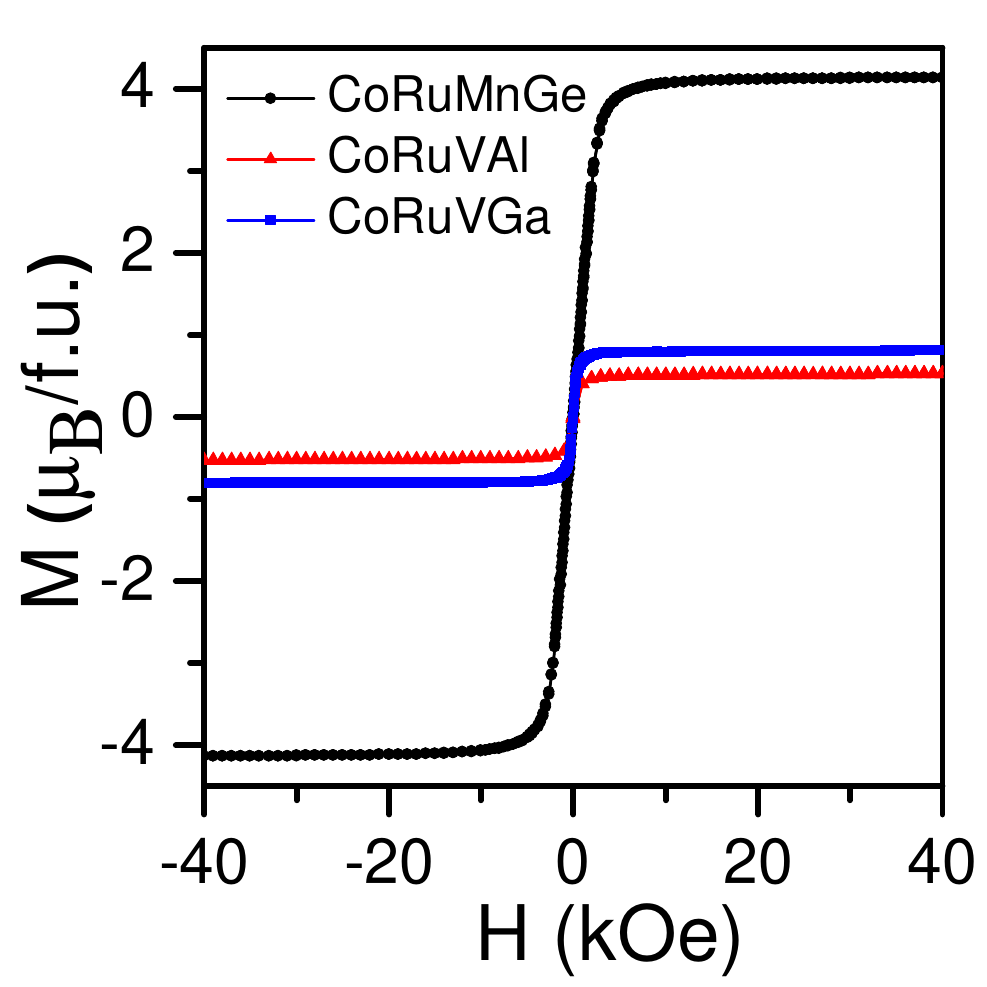}
	\caption{Isothermal magnetization curves at 5 K for CoRuMnGe, CoRuVAl and CoRuVGa alloys.}
	\label{MH}
\end{figure}
\begin{table}
	\centering
	\caption{Relaxed lattice parameter ($a_0$), atom-projected magnetic moments, total moments ($\mu_\mathrm{B}$) and relative energy ($\Delta E_{rel}$) of three non degenerate configurations Type I, II and III with respect to Type I for CRMG.}
	\begin{tabular}{l c c c c c c}
		\hline \hline
		& & & & & & \\
		Type& $\ $ $a_0$ (\AA) $ \ $  &  $m^{\mathrm{Co}}$ & $\ $ $m^{\mathrm{Ru}}$ $\ $  &  $m^{\mathrm{Mn}}$  & $\ $ $m^{\mathrm{Total}}$ $ \ $ & $\Delta E_{rel}$(eV/atom) \\ & & & & & &\\ \hline \\
		I    & 5.89   & 0.92 	&     0.09  	&	3.03 	& 4.03 		& 0.00   \\ & & & & & & \\
		II   & 5.81   & 0.75 	& 	 -0.08   	&   1.05	& 1.70		& 0.25  \\ & & & & & & \\
		III  & 5.85   & 1.23	& 	  0.87 		& 	0.89 	& 2.98 		& 0.33   \\ & & & & & & \\
		\hline \hline
	\end{tabular}
	\label{tab1}
\end{table}

\begin{table}
	\centering
	\caption{Relaxed lattice parameter ($a_0$), atom-projected magnetic moments, total moments ($\mu_\mathrm{B}$) and relative energy ($\Delta E_{rel}$) of three non degenerate configurations Type I, II and III with respect to Type I for CRVA.}
	\begin{tabular}{l c c c c c c}
		\hline \hline
		& & & & & & \\
		Type& $\ $ $a_0$ (\AA) $ \ $  &  $m^{\mathrm{Co}}$ & $\ $ $m^{\mathrm{Ru}}$ $\ $  &  $m^{\mathrm{V}}$  & $\ $ $m^{\mathrm{Total}}$ $ \ $ & $\Delta E_{rel}$(eV/atom) \\ & & & & & &\\ \hline \\
		I    & 5.88   & 0.72 	&     0.02  	&	0.28 	& 0.96 		& 0.00   \\ & & & & & & \\
		II   & 5.95   & 1.57 	& 	 0.40   	&   -0.32	& 1.67		& 0.38  \\ & & & & & & \\
		III  & 5.95   & 1.58	& 	  0.67 		& 	0.36 	& 2.63 		& 0.37   \\ & & & & & & \\
		\hline \hline
	\end{tabular}
	\label{tab2}
\end{table}

\begin{table}
	\centering
	\caption{Relaxed lattice parameter ($a_0$), atom-projected magnetic moments, total moments ($\mu_\mathrm{B}$) and relative energy ($\Delta E_{rel}$) of three non degenerate configurations Type I, II and III with respect to Type I for CRVG.}
	\begin{tabular}{l c c c c c c}
		\hline \hline
		& & & & & & \\
		Type& $\ $ $a_0$ (\AA) $ \ $  &  $m^{\mathrm{Co}}$ & $\ $ $m^{\mathrm{Ru}}$ $\ $  &  $m^{\mathrm{V}}$  & $\ $ $m^{\mathrm{Total}}$ $ \ $ & $\Delta E_{rel}$(eV/atom) \\ & & & & & &\\ \hline \\
		I    & 5.91   & 0.72 	&     0.02  	&	0.29 	& 0.99 		& 0.00   \\ & & & & & & \\
		II   & 5.95   & 1.57 	& 	  0.45   	&   -0.14	& 1.92		& 0.28  \\ & & & & & & \\
		III  & 5.96   & 1.53	& 	  0.73 		& 0.35 	& 2.64 		& 0.32   \\ & & & & & & \\
		\hline \hline
	\end{tabular}
	\label{tab3}
\end{table}

\begin{figure*}
	\centering
	\includegraphics[width=\linewidth]{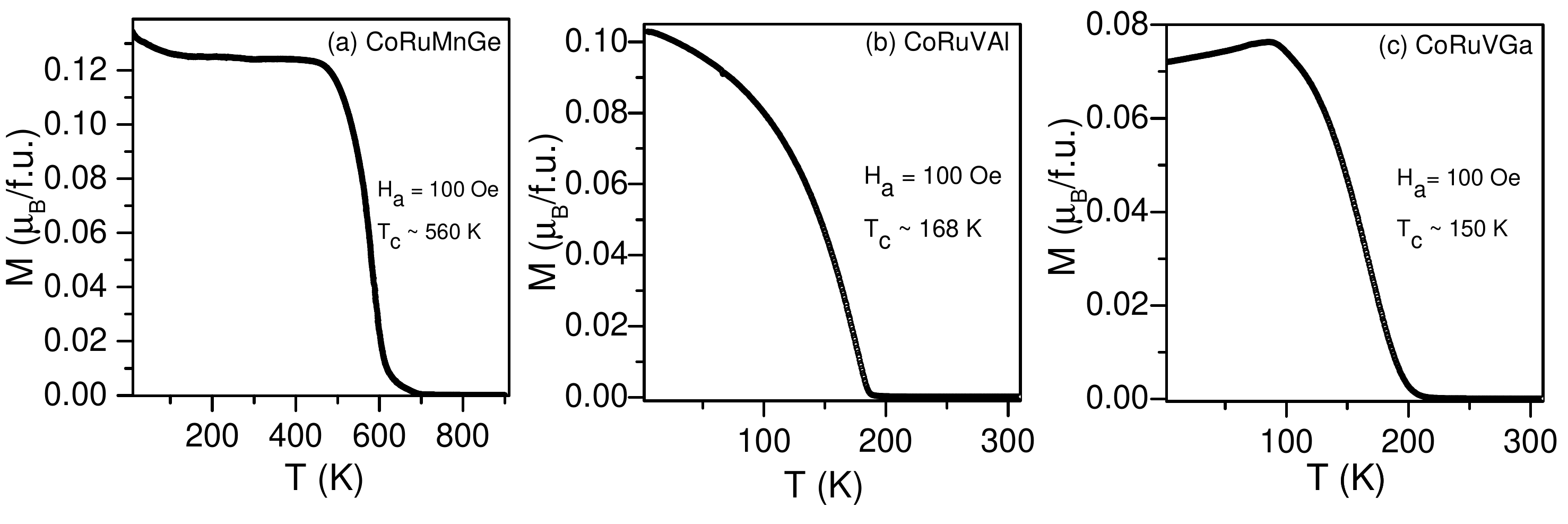}
	\caption{Magnetization (M) vs. temperature for (a) CoRuMnGe, (b) CoRuVAl and (c) CoRuVGa at 100 Oe. T$\mathrm{_C}$ is calculated by taking the minima of the first order derivative of M - T curve.}
	\label{MT}
\end{figure*}

\begin{figure}
	\centering
	\includegraphics[width=0.7\linewidth]{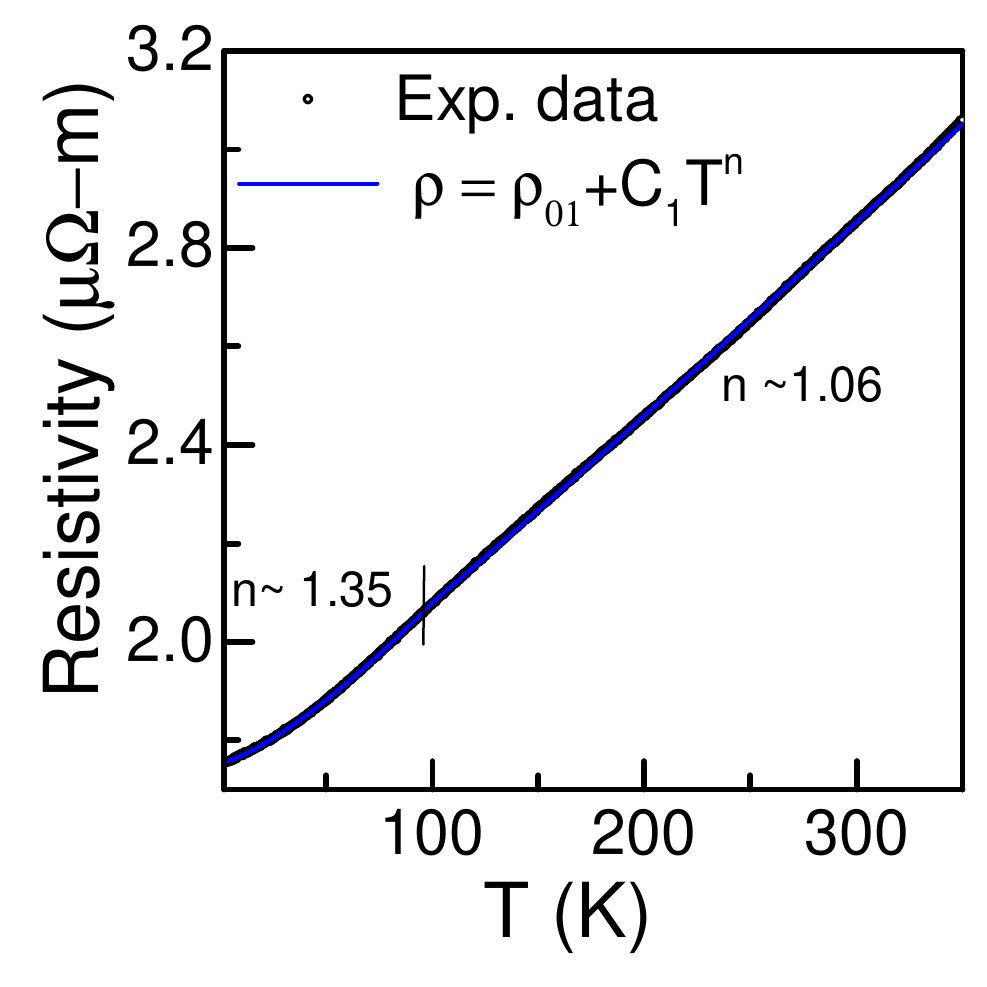}
	\caption{Temperature dependence of electrical resistivity($\rho$) for CoRuMnGe at zero field.}
	\label{res}
\end{figure}

\begin{figure}
	\centering
	\includegraphics[width=0.7\linewidth]{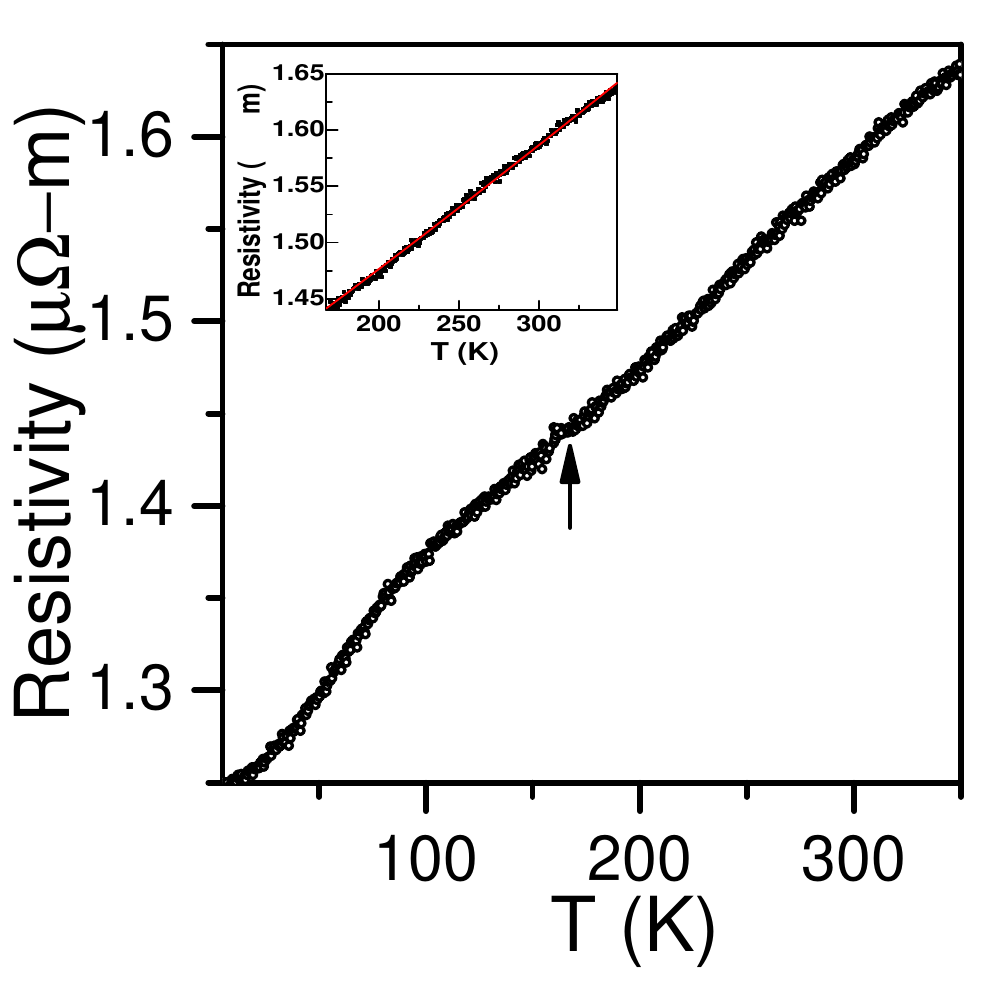}
	\caption{Temperature dependence of electrical resistivity($\rho$) for CoRuVAl at zero field.}
	\label{res1}
\end{figure}

\begin{figure}
	\centering
	\includegraphics[width=0.7\linewidth]{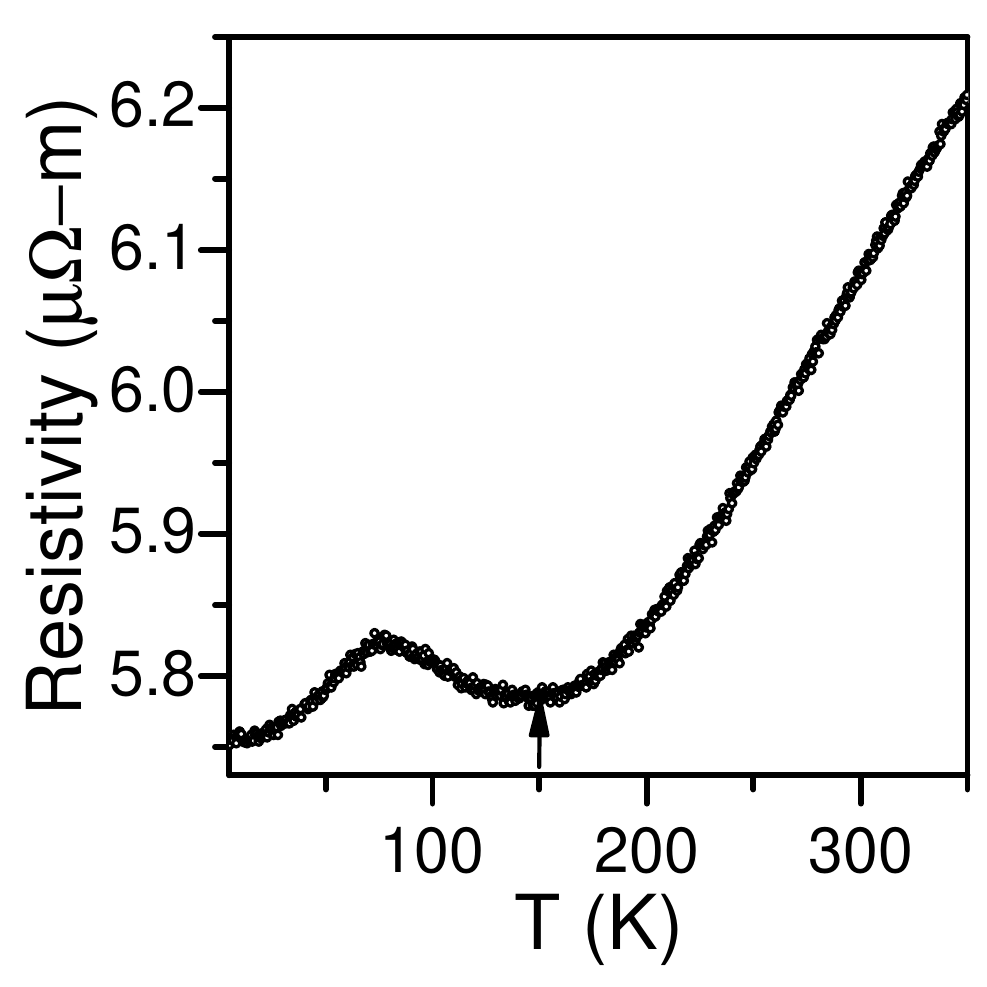}
	\caption{Temperature dependence of electrical resistivity($\rho$) for CoRuVGa at zero field.}
	\label{res2}
\end{figure}

Room temperature powder XRD patterns of CoRuMnSi, CoRuVAl and CoRuVGa alloys are shown in Fig. \ref{XRD}. From the patterns, it is clear that all the three alloys exhibit cubic crystal structure. Rietveld refinement of the XRD data was done using FullProf Suite software.\cite{RR} The lattice parameters as deduced from refinement were found to be 5.88, 5.90 and 5.91 $\mathrm{\AA}$ for CRMG, CRVA and CRVG respectively. The quaternary Heusler alloys exhibit LiMgPdSn-type structure whose primitive cell contains four atoms at the Wyckoff positions 4a, 4b, 4c and 4d. For the quaternary Heusler alloy considering X at 4b, $\mathrm{X^\prime}$ at 4c, Y at 4d and Z at 4a Wyckoff positions, the structure factor for the superlattice reflections (111) and (200) can be written as \cite{PhysRevB.99.104429}

\begin{equation}
	{F_{111}} = 4[{(f_Y-f_Z)-i(f_X-f_{X'})}]
\end{equation}
\begin{equation}
	{F_{200}} = 4[{(f_Y+f_Z)-(f_X-f_{X'})}]
\end{equation}
where, ${f_X, f_{X'}, f_Y}$ and ${f_Z}$ are the atomic scattering factors for X, X', Y and Z respectively.

\begin{figure*}
	\centering
	\includegraphics[width=\linewidth]{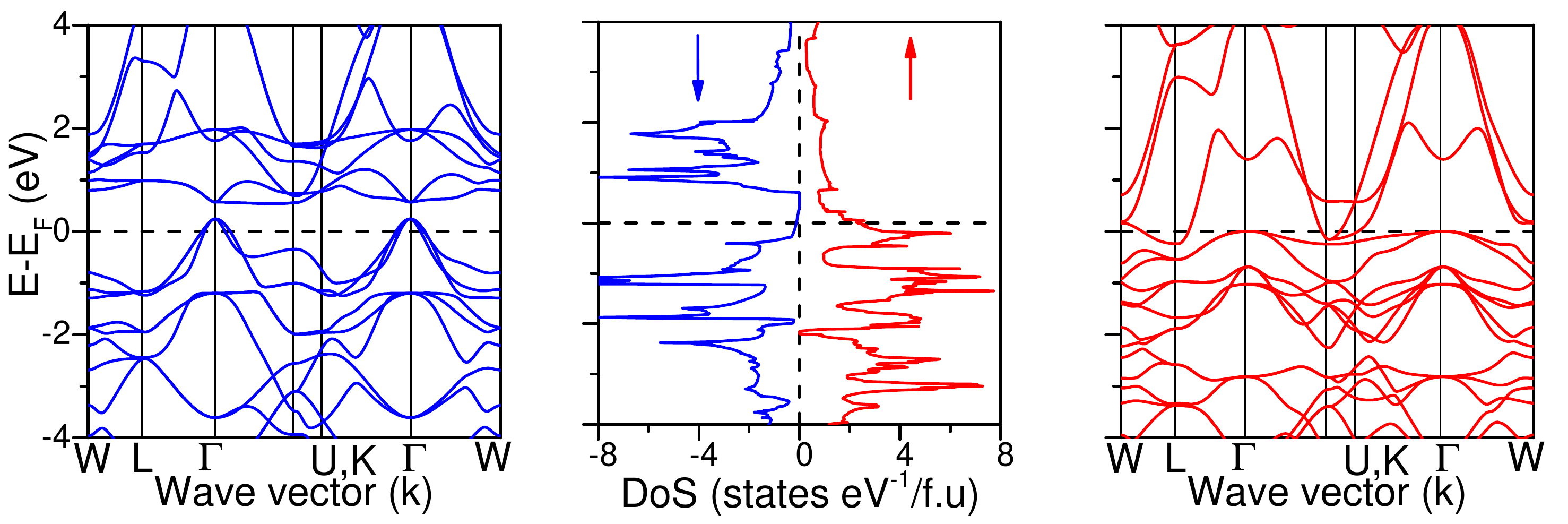}
	\caption{{Band Structure and the Density of states (DoS) for CRMG at experimental lattice parameter a$_{elp}$.}}
	\label{dos}
\end{figure*}

\begin{figure}
	\centering
	\includegraphics[width=0.7\linewidth]{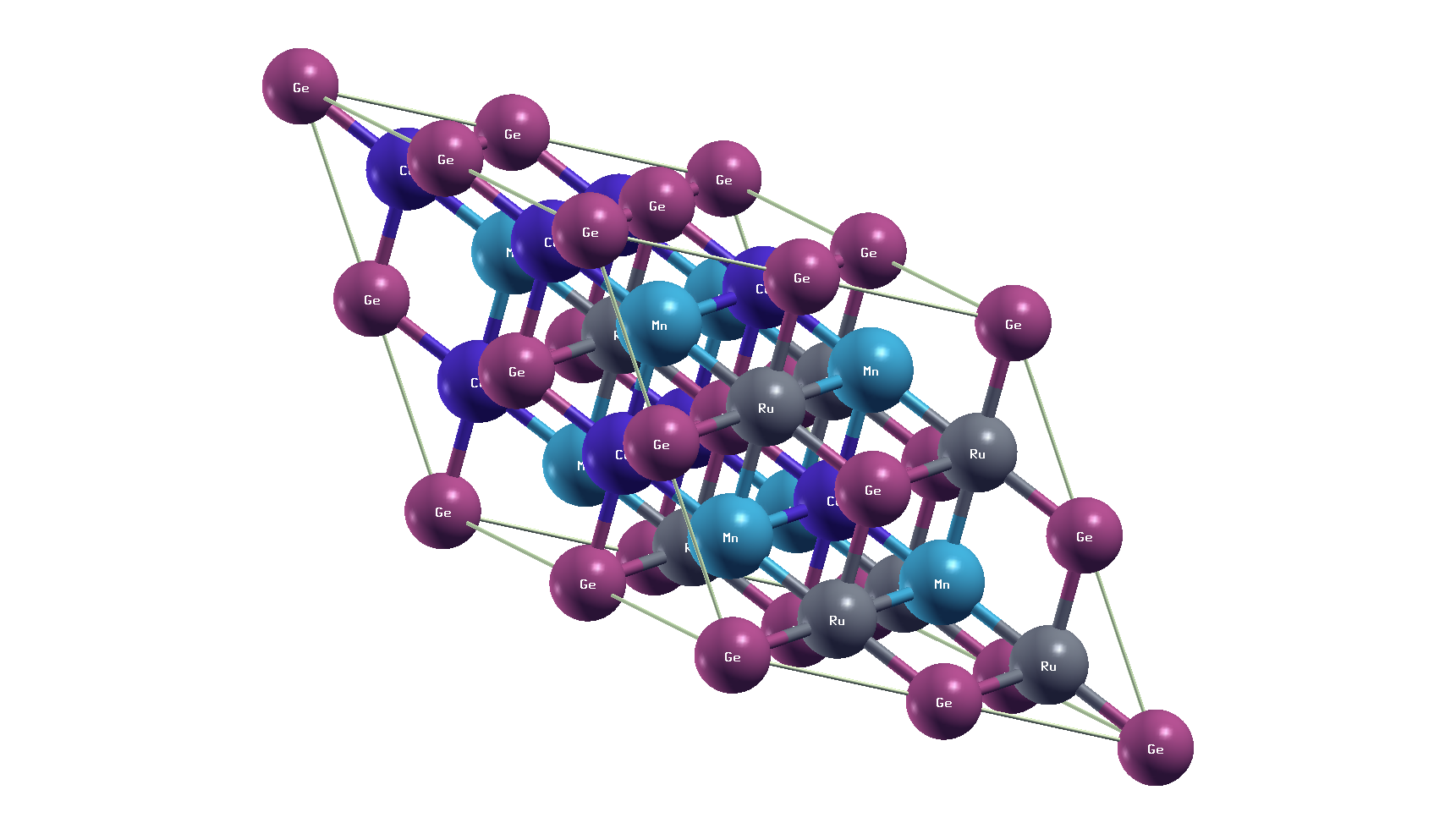}
	\caption{$2\times2\times2$ supercell of type I configuration of CoRuMnGe alloy. }
	\label{CRMG}
\end{figure}

\begin{figure}
	\centering
	\includegraphics[width=0.7\linewidth]{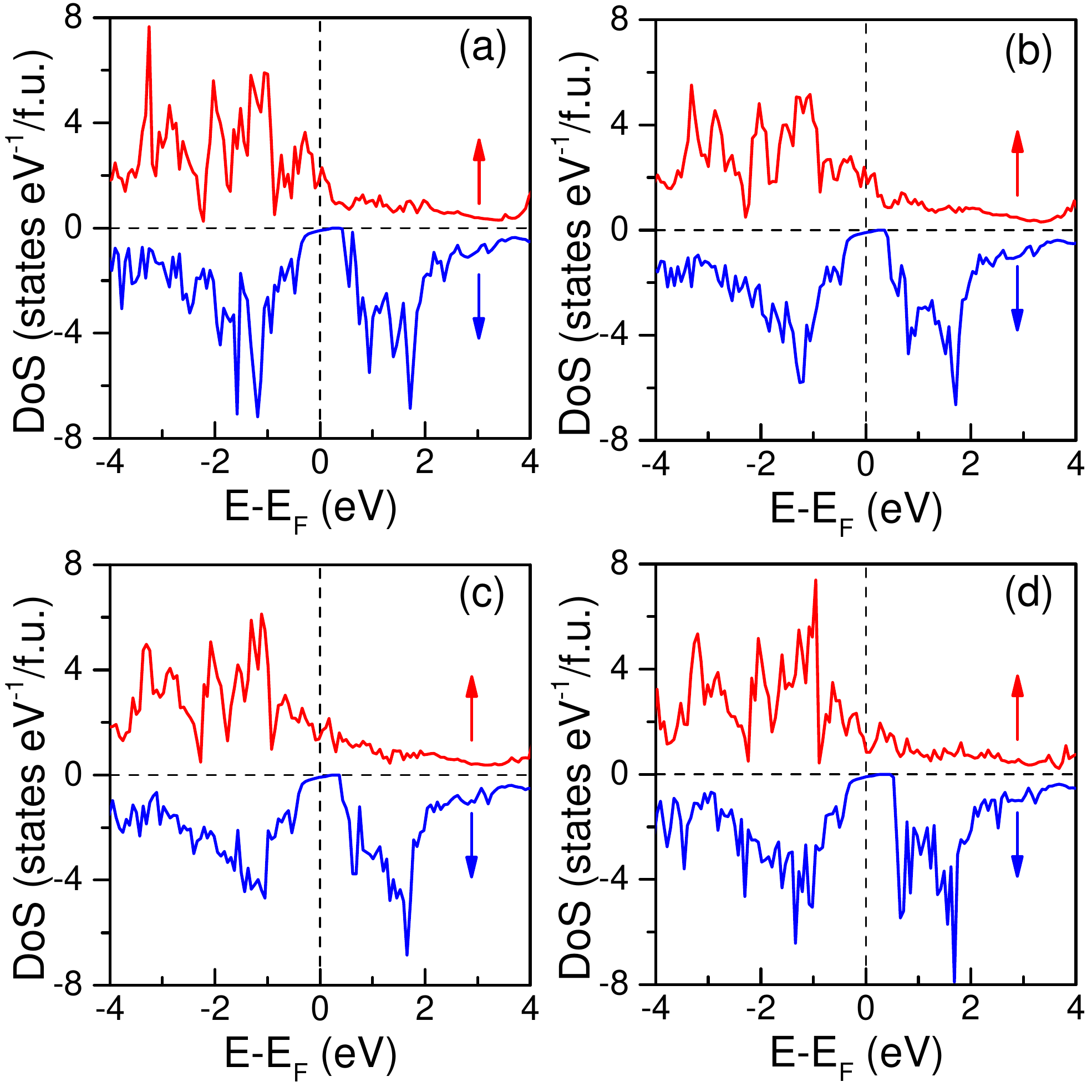}
	\caption{Density of states (DoS) for CoRuMnGe with (a) 12.5\%, (b) 25 \%, (c) 37.5 \% and (d) 50 \% Co-Ru swap disorder.}
	\label{disorder}
\end{figure}

\begin{figure*}
	\centering
	\includegraphics[width=\linewidth]{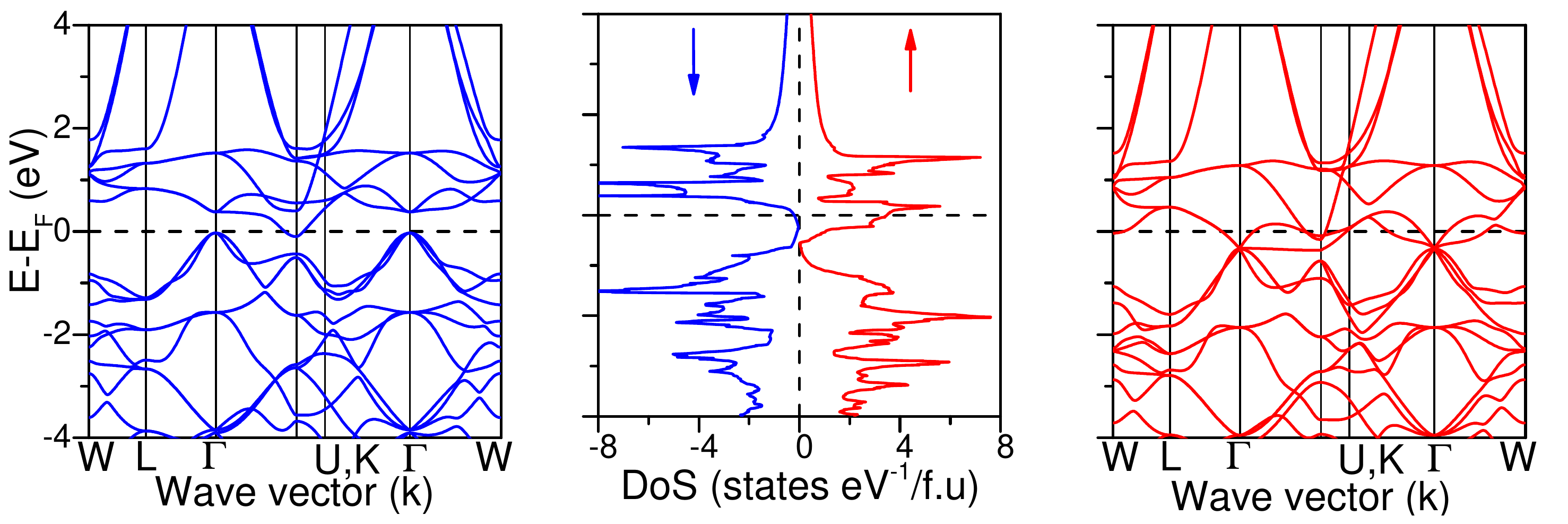}
	\caption{Band Structure and the Density of states (DoS) for CRVA at experimental lattice parameter (a$_{elp}$).}
	\label{CRVA_band}
\end{figure*}

\begin{figure*}
\centering
\includegraphics[width=\linewidth]{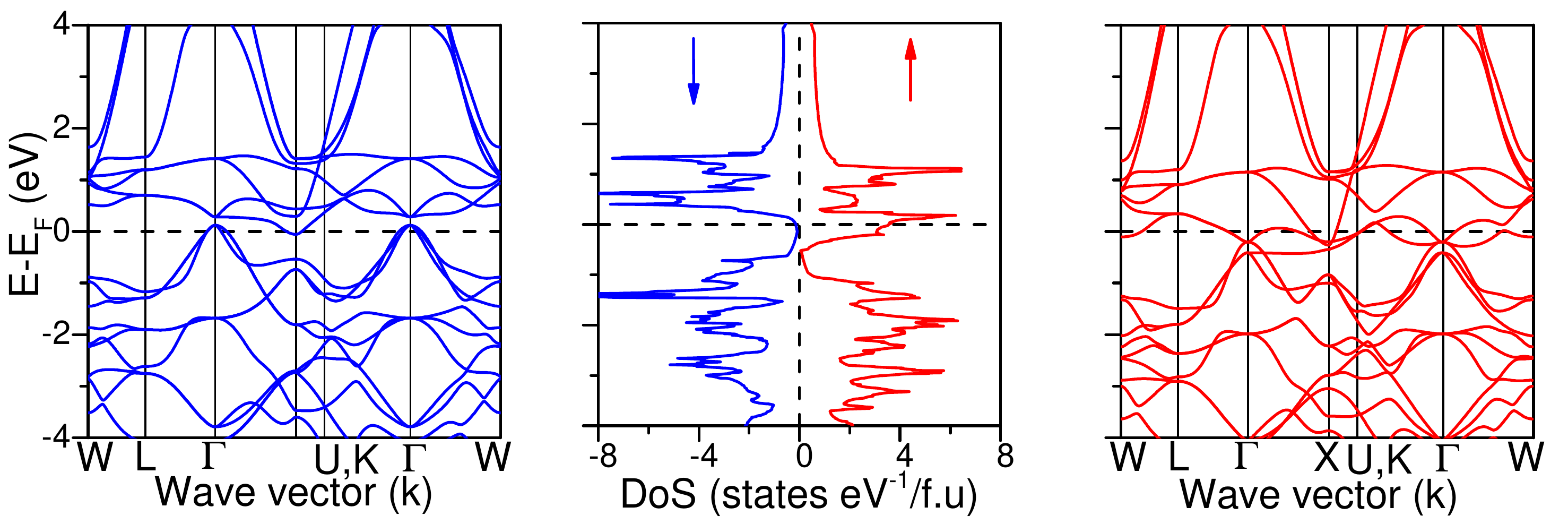}
\caption{Band Structure and the Density of states (DoS) for CRVG at experimental lattice parameter (a$_{elp})$.}
\label{CRVG_band}
\end{figure*}

As per the equation (1), in the case of B2 disorder (Y \& Z and X \& X$^\prime$ atoms are randomly distributed), the intensity of the (111) peak should be low or absent. On the other hand, for a completely disordered structure i.e., A2-type (all the four atoms occupy random positions), both the superlattice peaks (111) and (200) should be absent. Thus, to determine the disorder, the intensities of superlattice peaks (111) and (200) play a crucial role. The insets I of Fig. \ref{XRD} (a), Fig. \ref{XRD} (b) and Fig. \ref{XRD} (c) show the superlattice reflections with perfectly ordered LiMgPdSn-type structure for CRMG, CRVA and CRVG alloy respectively. It is clear that refinement considering ordered Y-type structure did not fit well. In case of CRMG, the XRD pattern fits well when disorder is considered between Co and Ru sites. Figure \ref{XRD} (a) shows the rietveld refined XRD pattern for CRMG with 50 \% disorder between Co and Ru sites i.e. the 4c and 4d sites are equally probable for Co and Ru atoms. Thus, due to 50\% swap disorder between tetrahedral site atoms (as revealed from refinement), the crystal symmetry reduces to L2$_1$. In case of CRVG and CRVA, the absence of (111) peak indicates complete B2 disorder. Figure \ref{XRD} (b) and \ref{XRD} (c) show the rietveld refined XRD pattern for CRVA and CRVG respectively with 50 \% disorder between Co \& Ru sites and V \& Al sites which results in complete B2 disorder. In this case, 4c \& 4d sites are equally probable for Co \& Ru atoms and 4a \& 4b sites are equally probable for V \& Al atoms. The insets II of Fig \ref{XRD}(b) and Fig \ref{XRD}(c) show the observed and calculated intensities of the superlattice reflections considering B2 disorder for CRVA and CRVG respectively. Thus, CRMG is found to have L2$_1$ structure, whereas CRVA and CRVG alloys tend to show B2-type disorder.

\begin{table*}
	\caption{Magnetic moments ($\mu_\mathrm{B}$) of ordered and disordered CoRuMnGe in a ($2\times2\times2$) supercell. $\mathrm{X_d}$ refers to defect atoms}
	\begin{tabular}{c| c c c |c |c |c c}
		\hline \hline
		& & & & & & \\
		System & Co & Ru & Mn & $\mathrm{X_d}$ & $\mathrm{m_{total}}$($\mu_B$) &$\mathrm{m_{total}}$($\mu_B$/f.u.) \\
		\hline\\
		Ordered & 0.91 & 0.08 & 3.03  & & 32.16 & 4.02\\
		\hline
		Co-Ru swap  & 0.96 & 0.03 & 3.07  & $\mathrm{Co_{Ru}}$: 1.02,& 32.10 & 4.01 \\
		(12.5\%)& 0.95 & 0.01 & 2.97  & $\mathrm{Ru_{Co}}$: -0.19 &\\
		& 0.97 & 0.04 & 3.09 & & & \\
		& 1.01 & 0.15 & 2.95 & \\
		&  & -0.03 & 3.08 & & \\
		\hline
		Co-Ru swap  & 1.00 & -0.01 & 3.08  & $\mathrm{Co_{Ru}}$: 1.05,1.05& 32.08 & 4.01\\
		(25.0\%)& 0.99 & -0.03 & 2.95 & $\mathrm{Ru_{Co}}$: -0.14,-0.15 &\\
		&0.98  & 0.08 & 3.04 & & \\
		& &-0.14  & 3.03 &  & \\
		\hline
		
		Co-Ru swap  & 0.97 & -0.07 & 3.04  & $\mathrm{Co_{Ru}}$: 1.04,1.04,1.05& 32.05& 4.01\\
		(37.5\%)& 1.01 & 0.00 & 3.01 & $\mathrm{Ru_{Co}}$: -0.10, -0.10, -0.11 &\\
		& 1.04 & 0.01 & 3.05  & & \\
		& & -0.04 & 2.96 & & \\
		& &  & 3.03 & & \\
		\hline
		Co-Ru swap  & 1.05 & -0.06 & 3.04  & $\mathrm{Co_{Ru}}$: 1.03,1.04,1.05,1.03& 32.07 & 4.01\\
		(50\%)& 1.04& -0.08 & 3.04 & $\mathrm{Ru_{Co}}$: -0.06,-0.06,-0.08,-0.06 &\\
		& 1.03 & -0.06 & 3.03 & & \\
		& &  & 3.01 & & \\
		
		\hline \hline
	\end{tabular}
	\label{MM}
\end{table*}

\begin{table}
	\centering
	\caption{Experimental lattice parameter ($a_{exp}$), Slater-Pauling, theoretically simulated and experimental total moments ($\mu_\mathrm{B}/f.u.$) and spin polarization (P) for the three alloys CRMG, CRVA and CRVG.}
	\begin{tabular}{l c c c c c c}
		\hline \hline
		& & & & & & \\
		Alloy&  $a_{exp}$    &  ${M_S}$(S-P) & ${M_S}$(Theor.)  &  ${M_S}$(Exp) & P (Theor.) &  \\
		&  $(\mathrm{\AA)}$    &  $({\mu_B}/f.u.$) & $({\mu_B}/f.u.$)  &  (${\mu_B}/f.u.$) & (\%) &  \\ & & & & & &\\ \hline \\
		CRMG    & 5.88   & 4.0 	&     4.03  	&	4.10 	& 91		  \\ & & & & & & \\
		CRVA   & 5.90   & 1.0 	& 	  0.96   	&   0.53	& 89		 \\ & & & & & & \\
		CRVG  & 5.91   & 1.0	& 	  0.99 		&   0.84 	& 93 	  \\ & & & & & & \\
		\hline \hline
	\end{tabular}
	\label{tab4}
\end{table}

\subsection{Magnetic Properties}
Figure \ref{MH} shows the variation of magnetization with field (M-H) at 5 K for CRMG, CRVA and CRVG. Quaternary Heusler alloys are known to follow the Slater-Pauling (S-P) rule according to which the total magnetic moment is directly related to the number of valence electrons in the unit cell as per the relation, \cite{PhysRevB.66.174429}
\begin{equation}
M = ( N_v - 24 )\hspace{0.1cm}\mu_B/f.u.
\end{equation}
CRMG has 28 valence electrons, whereas CRVA and CRVG have 25 valence electrons. Thus, as per the S-P rule CRMG, CRVA and CRVG are expected to have a magnetic moment of 4, 1 and 1 $\mu_B/f.u.$ respectively. From the M-H curves as shown in Fig. \ref{MH}, the calculated saturation magnetization values are found to be 4.10, 0.53 and 0.84 $\mu_B/f.u.$ for CRMG, CRVA and CRVG respectively. The small deviation in the magnetic moment from the S-P value in case of CRMG and CRVG alloys is due to the presence of small density of states in the minority spin channel at the Fermi level (See section {4.4} for details). This indicates a possibility of nearly half-metallic character in CRMG and CRVG alloys. A large deviation in the case of CRVA, may be attributed to the disorder. All the three alloys are found to be soft ferromagnetic with negligible hysteresis. \\

Figure \ref{MT} shows the thermo-magnetic (M-T) curves recorded under an applied field of 100 Oe. The Curie temperature, T$\mathrm{_C}$ is estimated by taking the minima of the $\frac{dM}{dT}$ vs. T curves. T$\mathrm{_C}$ for CRMG, CRVA and CRVG is found to be 560, 168 and 150 K respectively. A small increase in the magnetic moment in the lower temperature range is possibly due to anisotropic variations with temperature, since the M vs. T curves are measured in low fields of 100 Oe. Among the three alloys, CRMG is found to have the highest T$\mathrm{_C}$ and thus is suitable for room temperature spintronic applications. 

\subsection{Transport properties}
\textbf{1. CoRuMnGe}\\
Figure \ref{res} shows the temperature dependence of electrical resistivity ($\rho_{xx}$) for CRMG in the temperature range 5 - 350 K at zero field. The resistivity increases with temperature indicating the metallic nature. To further investigate the $\rho_{xx}$ vs. T behaviour, the resistivity curve was fitted using the power law given by:
\begin{equation}
\rho(T) = \rho_{0}+ \rho(T) = \rho_{0} + BT^n
\end{equation}
In previous reports on half-metallic Heusler compounds, different values of n are reported depending on the temperature range considered.\cite{PhysRevB.96.184404,0022-3727-40-6-S01, 0022-3727-37-15-001, doi:10.1063/1.126606, PhysRevApplied.10.054022} 
In the temperature region $\mathrm{35 K < T < 100 K}$ (Region I), the value of n was found to be 1.35 whereas, for T $>$ 100 K (Region II), the resistivity varies almost linearly with n = 1.06. The linear dependence in the region II can be attributed to the electron-phonon scattering. In a half-metal, there are no minority spin charge carriers at E$_{F}$ due to complete spin polarization and thus, spin-flip scattering is usually not possible.\cite{1989JPCM1.2351O,Kubo-1972} Due to this, the $\mathrm{T^2}$ term related to single magnon scattering is expected to be absent in resistivity. From the value of n, we can conclude that the dependence is not quadratic which indirectly confirms the half-metallic nature in CRMG alloy. Also, the residual resistivity and the residual resistivity ratio (RRR = ${\rho_{300 K} / \rho_{5 K}}$) values are found to be 1.75 $\mathrm{\mu \Omega}$ m and 1.63 respectively. The obtained RRR value is highest among the three alloys under study, indicating the possibility of least disorder.
\\

\textbf{2. CoRuVAl}\\
Figure \ref{res1} shows the temperature dependence of electrical resistivity ($\rho_{xx}$) for CRVA in the temperature range of 5 - 350 K at zero field (the arrow indicates the Curie temperature). The alloy shows metallic nature in both ferromagnetic and paramagnetic regions. It should be noted that below T$\mathrm{_C}$, the change in resistivity is mainly due to magnetic scattering, whereas in the paramagnetic state the magnetic component of resistivity saturates. Thus, for $\mathrm{T > T_C}$ (169 K), the change in resistivity with temperature is determined by electron-phonon scattering only, which is usually observed as a linear contribution in $\rho$(T) above the Debye temperature. Inset of Fig. \ref{res1} shows the linear behaviour of $\rho$ vs. T for $\mathrm{T > T_C}$. The residual resistivity is found to be 1.25 $\mathrm{\mu \Omega}$ m and the RRR value is found to be 1.27. 
\\

\textbf{3. CoRuVGa}\\
Figure \ref{res2} shows the temperature dependence of resistivity ($\rho_{xx}$) for CRVG in the temperature range of 5 - 350 K at zero field. In this case, the resistivity behavior is significantly different and shows unconventional features like: (a) large value of residual resistivity ($\rho_0$ = 575 $\mathrm{\mu \Omega}$ cm), (b) anomaly in the form of a maxima below T$\mathrm{_C}$, and (c) presence of region with semiconducting behavior i.e., with a negative coefficient of resistivity in the ferromagnetic state. The two main factors which determine the resistivity behavior in Heusler alloys are (1) conduction electron scattering mechanism and (2) effect of magnetic ordering on the electronic band structure near the Fermi level.  In the low temperature region ($\mathrm{T < 80 }$ K), the resistivity decreases with decreasing temperature and hence shows the metallic nature. This decrease in resistivity can be attributed to the reduction in the magnetic scattering with decreasing temperature. Anomaly in the form of a maximum seen in resistivity in the magnetically ordered state is not something new. It has been observed in a few systems that, the effect of change in electronic structure strongly reflects in the temperature dependence of resistivity. This causes an anomaly in the form of a maximum because of the superposition of the electron-phonon and the magnetic contributions, when an abrupt decrease in magnetic contribution induced by vanishing spontaneous magnetization is superimposed by a linear increase in phonon contribution.\cite{DC} The electron-transport behavior in half-metallic ferromagnetic Heusler alloy Co$_2$CrGa was found to show a similar behavior as in our system CRVG. \cite{Kourov2013, KOUROV2015839}  Below the Curie temperature, the change in resistivity is mainly caused by the magnetic contributions. In case of half-metallic ferromagnets, the magnetic contribution to the conductivity is determined by considering two parallel conduction channels for electrons, one with spin-up and the other with spin down and the total magnetic contribution of conductivity can be written as, \cite{Kourov20131}
\begin{equation}
\sigma_m = \sigma_{\downarrow} + \sigma_{\uparrow} \hspace{0.5cm}\mathrm{or} \hspace{0.5cm} \rho_m = \frac{\rho_{\downarrow}  \rho_{\uparrow}} {\rho_{\downarrow} + \rho_{\uparrow}}
\end{equation}
The conductivity of spin-up electrons is determined mainly by the scattering of the charge carriers. For the spin down electrons, the conductivity is dependent on the energy gap parameters in the electronic spectrum. 
Also, the energy gap parameters mainly depend on the spontaneous magnetization ($\mathrm{M_S}$).\cite{DC,Kourov20131} At $\mathrm{T \ll T_C}$, $\mathrm{M_S}$ does not vary much with temperature and thus, the energy gap parameters and conductivity of spin down channel remain almost constant. Due to this, at low temperature, the variation in resistivity is mainly due to conduction of spin up electrons. At T $\rightarrow$ $\mathrm{T_C}$, the saturation magnetization vanishes which results in the disappearance of energy gap in the spin down channel and hence increase in its conductivity ($\mathrm{\sigma_{\downarrow}}$). As a result, $\mathrm{\rho_m}$ decreases. And thus, we observe a region of negative TCR near the Curie temperature. A negative TCR value has been reported in materials like V and Ti doped $\mathrm{Fe_3Ga}$, \cite{PhysRevB.44.12406} Ti, Mn, Cr doped $\mathrm{Fe_3Si}$ \cite{PhysRevB.48.13607} and was speculated to arise due to the existence of small electronic density at $\mathrm{E_F}$ and large spin disorder scattering. Similar behavior is also reported in various half-metallic ferromagnetic Heusler alloys where a region of negative TCR is observed and was explained on the basis of presence of a gap in the electronic spectrum near E$_F$.\cite{Kourov2013, PhysRevB.72.012417, Kourov2016, MARCHENKOV2018211} It has also been observed that a negative TCR is generally seen in alloys when the electrical resistivity $\rho > 1.5 \mu \Omega$ m in the paramagnetic state.\cite{PhysRevB.44.12406, PhysRevB.48.13607} For CoRuVGa, the resistivity in the paramagnetic state lies in the range of 5.79 to 6.20 $\mathrm{\mu \Omega}$ m and thus it fulfills this condition to show anomaly in the resistivity near the Curie temperature.  In the paramagnetic state [$\mathrm{T > T_C}$ (150 K), the resistivity increases almost linearly with temperature which can be attributed to the electron-phonon scattering since in this region $\mathrm{\rho_m}$ is constant. The residual resistivity value is found to be 5.75 $\mathrm{\mu \Omega}$ m, which is the highest among the three alloys under study. The RRR is found to be  low (1.06) as compared to other two alloys. 

\subsection{Theoretical Results}
The magnetic configuration of the alloys was examined by simulating different initial magnetic (para-, ferro- and ferrimagnetic) arrangements. The results of the structural optimization for CRMG, CRVA and CRVG are displayed in Table \ref{tab1}, \ref{tab2} and \ref{tab3} respectively. It showed that all the three alloys are stable in configuration I with ferromagnetic ordering as it exhibits the lowest total energy. The most stable, (Type I) configuration has Co at 4d (0.75,0.75,0.75), Ru at 4c (0.25,0.25,0.25), Y (Mn,V) at 4b (0.5,0.5,0.5) and Z (Ge,Ga,Al) at 4a (0,0,0) sites. It is seen that, in case of CRMG, most of the magnetic moment is contributed by Mn (3 $\mu_B$), whereas Co contributes a small moment (1 $\mu_B$) and Ru has negligible moment. In case of CRVA and CRVG, moments are mainly contributed by Co and V, whereas Ru has negligible moment. To further study the electronic structure and magnetic properties of these alloys, the energetically most favourable Type I structure was used. Figure \ref{dos} shows the calculated spin resolved band structure and density of states (DoS) for CRMG alloy, calculated at the experimental lattice parameter ($a_{elp}$). The alloy manifests electronic structure of a highly spin-polarized, nearly half-metal. The ideal half metal acquires 100 \% spin polarization. In this case, due to the presence of small number of states (0.11 states/eV/f.u.) at E$\mathrm{_F}$ for the minority spin channel, the alloy has slightly low spin polarization(P = 91 \%). The calculated moment is 4.03 $\mu_B/f.u.$, which is close to that of Slater-Pauling value (4 $\mu_B/f.u.$).

Going back to our XRD results, it was found that 4c and 4d fcc sites in CoRuMnGe are equally possible for Co and Ru atoms (see section 4.1).  In order to get a deeper insight into the properties of this actual experimental structure, one should simulate a mixture of Co and Ru atoms at these two sites. One way to do this is to swap the position of Co and Ru atoms at 4c and 4d sites. To simulate such intrinsic disorder, a $2\times2\times2$ supercell of the primitive cell of the type I configuration of CRMG is constructed (See Fig. \ref{CRMG}). In a $2\times2\times2$ supercell with 32 atoms (eight formula units), 12.5\% swap disorder was simulated by exchanging one of the eight Co atom positions and one of the eight Ru atom positions. Similarly, 50\% swap disorders were simulated by exchanging four of the eight Co atoms and four of the eight Ru atom positions. All possible configurations for replacement of Co by Ru and vice versa was checked, and energetically the most stable configuration is chosen to present the result here. The Co-Ru swaps almost give the same total magnetic moment as the ideal (no swap) case. The calculated DoS for disordered structures are shown in Fig. \ref{disorder}. 
As seen from Fig. \ref{disorder}, half-metallicity in CoRuMnGe is quite robust against swapping disorder between Co ad Ru sites. In the ordered Y-type structure, Co and Ru atoms are surrounded by four Al and 4 Mn atoms as their nearest neighbors. On considering 50 \% swap disorder between Co and Ru sites, which results in the disordered $L2_1$ type structure, it is seen that the local environment of Co and Ru atoms remains the same i.e. they are still surrounded by four Al and 4 Mn atoms, however there is a slight change in the bond length. It turns out that, with swap disorder, the local environment of the defected sites does not affect the exchange interaction much but causes a little change in the bond length (due to relaxation effect). This, in turn, changes the local moments (and hence the total cell moment) by a little amount only. Thus, in case of CoRuMnGe, the swap disorder between Co and Ru sites does not change the total magnetic moment as well as the density of states. To have a better understanding, we have calculated the local moments at the individual atomic sites, the values of which are shown in Table \ref{MM}. Here $\mathrm{X_d}$ refers to the sites where Co and Ru atoms are swapped. In each case, the magnetic moment of Ru swapped with Co atom is changing from positive to negative (making it antiferromagnetically aligned) as compared to the ideal case,however Ru has negligible moment in both case. Also, the magnetic moment of Co swapped with Ru atoms slightly increases. Co-Ru swap defected structures almost maintain the same total magnetic moment as the ideal structure i.e. $\sim$ 4.0 $\mu_B$/f.u.\\

The DoS plots for CRVA and CRVG alloys are shown in Fig. \ref{CRVA_band} and Fig. \ref{CRVG_band} respectively, which reveal a nearly  half-metallic character. The spin polarization is found to be 89 and 93 \% for CRVA and CRVG respectively. In the minority spin channel, a small number of states ($\mathrm{N_{\downarrow}}$ = 0.20 states/eV/f.u. for CRVA and 0.13 sates/eV/f.u. for CRVG) are present at $\mathrm{E_F}$, which is responsible for a slightly lower spin polarization as compared to the ideal half-metal. The magnetic moment is close to the expected value as per Slater-Pauling rule. To have a one-to-one comparison, the experimental lattice parameters, theoretical, simulated and experimental magnetic moments and simulated spin polarizations are tabulated in Table \ref{tab4}. It is seen that the experimentally observed moment for CRVA is quite different from the calculated value. To understand the reason behind such difference, we have performed swap disorder and anti-site disorder calculations for CRVA, details of which are provided in the supplementary material.\cite{RR3} The presence of such disorder is justified in CRVA because of the B2-type disorder as predicted from XRD data. As in the case of CoRuMnGe, the swap disorder does not alter the net magnetic moment for CRVA as well and the total moment still remains close to 1.0 $\mu_B /f.u.$. The total magnetic moment, however, are extremely sensitive to the anti-site disorder (antisite between Co \& Ru and V \& Al). Interestingly, due to the change in the local environment in this case, the local atomic moments change drastically; sometimes even get quenched and/or antiferromagnetically aligned as compared to the completely ordered case. The latter is attributed to the itinerant character of magnetism in Co, Ru, and V-containing Heusler alloys, and to the frustration of antiferromagnetic exchange interactions, possibly accompanied by a tetragonal distortion. In fact, the exchange interaction in this case is reasonably long ranged, affecting the moment of atoms sitting far from the defected sites. Table S2 of the supplement file shows the simulated results for a $2\times2\times2$ supercell which reduces the net cell moment up to 0.57 $\mu_B /f.u.$, when antisite disorder between both (Co \& Ru) and (V \& Al) pairs are considered. Similar behavior has been reported in other Heusler alloys, where swap/anti-site disorder causes a decrease in moment.\cite{doi:10.1063/1.4972797, JOUR, doi:10.1063/1.4998308} When comparing CoRuVGa and CoRuVAl, the electronegativity value of Al (1.61) is similar to that of V(1.63). Due to this, there is a high probability of Al atom to occupy one of the octahedral sites (1/2, 1/2, 1/2) i.e. the disorder between V and Al is much more probable to occur due to their similar electronegativity values. This, however, is not the case for Ga (1.81) and V(1.63) in CoRuVGa. This is probably the reason for a stronger swap/antisite disorder in CRVA as compared to CRVG and hence a larger reduction in the moment in the former as compared to later.
\section{Conclusion}
In conclusion, we have performed a detailed experimental and theoretical study on the structural, magnetic, transport and electronic properties of CoRu- based quaternary Heusler alloys CoRuMnGe, CoRuVAl and CoRuVGa. All the three alloy were found to crystallize in cubic structure. In the case of CRMG, the XRD analysis reveals that the tetrahedral sites (Co and Ru) are equally probable, due to which its crystal symmetry is reduced to L2$_1$. On the other hand, for the other two alloys CRVA and CRVG, the absence of (111) superlattice reflection gives the indication of B2-type disorder. For CRMG and CRVG, the experimental magnetic moment values are in good agreement with the theory as well as Slater-Pauling rule and for CRVA, a relatively large deviation is found. The reduction in the magnetic moment in case of CoRuVAl possibly arises due to the anti-site disorder between Co \& Ru sites and V \& Al sites. The resistivity measurements indirectly support the half-metallic behavior in CRMG alloy. The highest T$\mathrm{_C}$ of 560 K was found for CRMG alloy. CRVA was found to be metallic in the ferromagnetic as well as paramagnetic state. A strong dependence of magnetic and electronic structure is seen in the temperature dependence of resistivity for CRVG which shows a maximum and a region of negative TCR value in the ferromagnetic region. The ab intio calculations predict the nearly half-metallic ferromagnetic state with high spin polarization in all these alloys. Thus, nearly half-metallic character, high T$\mathrm{_C}$ and high spin polarization makes CoRuMnGe alloy promising for room temperature spintronic applications.

\section*{Acknowledgment}
DR thank Council of Scientific and Industrial Research (CSIR), India for providing Senior Research Fellowship. AA  acknowledge National Center for Photovoltaic Research and Education (NCPRE), IIT Bombay, India for possible funding to support this research.

\section*{References}
\bibliography{bib}

\end{document}